\documentstyle[prd,aps,floats,12pt]{revtex}
\tightenlines

\def\lsim{\
  \lower-1.5pt\vbox{\hbox{\rlap{$<$}\lower5.3pt\vbox{\hbox{$\sim$}}}}\ }
\def\gsim{\
  \lower-1.5pt\vbox{\hbox{\rlap{$>$}\lower5.3pt\vbox{\hbox{$\sim$}}}}\ }

\begin{document}

\title{Linearized gravity on the Randall--Sundrum two-brane background
with curvature terms in the action for the branes}

\author{Yuri Shtanov$^a$\footnote{E-mail: shtanov@bitp.kiev.ua}
and Alexander Viznyuk$^{a,\,b}$\footnote{E-mail: viznyuk@bitp.kiev.ua}}
\address{$^a$Bogolyubov Institute for Theoretical Physics, Kiev 03143,
Ukraine\\ $^b$Department of Physics, Taras Shevchenko National University, Kiev
03022, Ukraine}

\maketitle

\vspace{1cm}

\begin{abstract}
We study gravitational perturbations in the Randall--Sundrum two-brane
background with scalar-curvature terms in the action for the branes, allowing
for positive as well as negative bulk gravitational constant. In the zero-mode
approximation, we derive the linearized gravitational equations, which have the
same form as in the original Randall--Sundrum model but with different
expressions for the effective physical constants.  We develop a generic method
for finding tachyonic modes in the theory, which, in the model under
consideration, may exist only if the bulk gravitational constant is negative.
In this case, if both brane gravitational constants are nonzero, the theory
contains one or two tachyonic mass eigenvalues in the gravitational sector. If
one of the brane gravitational constants is set to zero, then either a single
tachyonic mass eigenvalue is present or tachyonic modes are totally absent
depending on the relation between the nonzero brane gravitational constant and
brane separation. In the case of negative bulk gravitational constant, the
massive gravitational modes have ghost-like character, while the massless
gravitational mode is not a ghost in the case where tachyons are absent.
\end{abstract}


\section{Introduction}

Linear gravitational perturbations of the flat braneworld models were studied
beginning from the seminal papers by Randall \& Sundrum \cite{RS1,RS2}, where
their spectrum was shown to contain, besides the zero mode, also an infinite
tower of Kaluza--Klein massive modes.  Since then, perturbations in various
types of braneworld scenarios and in various approximations were considered in
the flat case as well as on the cosmological background (which is a much more
complicated issue still far from being well understood; see \cite{cosmology}
and references therein). In this paper, we study certain aspects of linear
perturbations on a particular simple background, which, to our knowledge, have
not been previously investigated.  We consider the Randall--Sundrum two-brane
model (the so-called RS1 model \cite{RS1}) supplemented by scalar-curvature
terms in the action for both branes.  Historically, the induced
scalar-curvature term for the brane was introduced in \cite{DGP} (see also
\cite{CHS}) as a method of making gravity on the brane effectively
four-dimensional even in the flat infinite bulk space.  The corresponding
cosmological models were initiated in \cite{CHS,DDG}. The effects of these
scalar-curvature terms in the two-brane setup under consideration were recently
studied in \cite{effects,CD,Padilla,Smolyakov,mirror}. Perturbations in the
one-brane (RS2 \cite{RS2}) counterpart of this model were previously
investigated in \cite{KTT,Tanaka}.

The study of the braneworld models is usually confined to the case of positive
bulk and brane gravitational constants. (The signs of the gravitational
constants in this case are defined relative to the signs of the conventional
matter Lagrangians on the branes.) This can be explained by the fact that only
positive values of the bulk gravitational constant are allowed in the
Randall--Sundrum model, since its sign coincides with the sign of the effective
Newton's constant in that model \cite{RS1,RS2}. When one adds curvature terms
in the action for the branes, this assumption might be relaxed, and, in this
paper, we allow for positive as well as negative bulk gravitational constant
while keeping gravitational constants on the branes positive. More generally,
one could consider various relations between the signs of the four-dimensional
gravitational constants in the action for the branes, bulk gravitational
constant, and conventional matter Lagrangians
\cite{CD,Padilla,Smolyakov,mirror}. Although we are not aware of any
fundamental multidimensional theory that can produce different signs of the
gravitational couplings for the bulk and branes, we consider this possibility
from the viewpoint of the effective action regardless of the unknown underlying
theory.

The study of both signs of the bulk gravitational constant is partially
motivated by the existence of braneworld models with interesting behaviour that
require negative brane tension. One of them is  the model of {\em disappearing
dark energy\/} (DDE) \cite{SS,AS}. The DDE model is a braneworld model of
expanding universe which, after the current period of acceleration, re-enters
the matter-dominated regime continuing indefinitely in the future. The merit of
this model of dark energy is the absence of the cosmological event horizon
owing to the fact that the universe becomes flat, rather than De~Sitter, in the
asymptotic future. The DDE model is based on the generic braneworld action with
the bulk cosmological constant and brane tension satisfying the
Randall--Sundrum constraint, and also including the curvature term in the
action for the brane. For the consistency of this model with the current
cosmological observations (specifically, for the condition $\Omega_{\rm m} < 1$
on the dark-matter cosmological parameter), the brane tension has to be {\em
negative\/} \cite{SS}.  Negative brane tension is required also for the
existence of unusual `quiescent' singularities \cite{SS1} in the AdS-embedded
braneworld models, which occur during the universe expansion and are
characterized by {\em finiteness\/} of the scale factor, Hubble parameter, and
matter density.

The bulk gravitational constant enters the homogeneous cosmological equations
on the brane in even power; therefore, its sign does not matter on the level of
the homogeneous cosmology on the brane \cite{SS}. However, the relation between
the sign of the bulk gravitational constant and the sign of the brane tension
is of importance for the small-scale gravitational physics in the braneworld,
in particular, for the behaviour of cosmological perturbations. This can be
seen already from the fact that the property of `localization' of
five-dimensional gravity in the neighbourhood of the brane (that the warp
factor locally decreases as one moves away from the brane) requires the brane
tension and the bulk gravitational constant to be of the same sign.  This
localization property may turn out to be important for a consistent braneworld
theory and, therefore, for negative-tension branes, it may require negative
gravitational constant in the bulk. Thus, it seems important to keep open this
possibility when generalizing the Randall--Sundrum model by including the
curvature terms in the action for the branes.

The issue of ghosts in a theory with positive bulk gravitational constant but
arbitrary signs of the {\em brane\/} gravitational constants was recently under
investigation in \cite{Padilla}, and certain regions of parameters were ruled
out.  In this paper, we consider somewhat complimentary situation where the
brane gravitational constants are always positive with respect to the
conventional matter Lagrangian, while the bulk gravitational constant can be of
any sign.\footnote{The results obtained in one of such theories are usually
easy to apply to another theory by using the change of the overall sign of the
action.} We will be mainly concerned with the issue of tachyons in the theory
of this kind, which, to our knowledge, was not discussed in the literature
before. Our original method of finding tachyonic modes in the two-brane
background is quite general and can be applied to other braneworld models
leading to the same system of equations for the bulk gravitational modes, in
particular, to the model with the Gauss--Bonnet action in the bulk considered
in \cite{CD}.

It will be shown below that, unlike in the pure Randall--Sundrum case, the
presence of the curvature terms  in the action for the branes leads to a
possibility of unwanted tachyonic modes if the bulk gravitational constant is
negative.  We demonstrate that, in this case, there can be only one or two
tachyonic mass eigenvalues in the theory under consideration and determine the
range of parameters for which tachyonic modes do not exist. We will also see
that the massive gravitational modes have ghost-like character in the case of
negative bulk gravitational constant, while the massless gravitational mode is
not a ghost in the case where tachyons are absent.  These results are in
agreement with those of \cite{Padilla}. Thus, it is mainly the presence of
ghosts in the Kaluza--Klein massive spectrum of gravity that makes models with
negative bulk gravitational constant problematic.

This paper is organized in the following manner.  After describing the model,
we review a suitable theory of linear gravitational perturbations on the
Randall--Sundrum two-brane background. Then, in the zero-mode approximation, we
derive the linearized system of gravitational equations with matter confined to
the visible and/or hidden brane.  These equations have the same form as in the
original RS1 model but with different physical constants.  After that, we
specially investigate the case of negative bulk gravitational constant and show
that the linearized theory can contain tachyonic gravitational modes. In this
case, one or two tachyonic mass eigenvalues are observed if both brane
gravitational constants are nonzero.  If one of the brane gravitational
constants is zero, then either a single tachyonic mass eigenvalue is present or
tachyonic modes are totally absent; which of these two possibility is realized
depends upon the values of the nonzero brane gravitational constant and brane
separation. We determine the range of parameters for which tachyonic modes are
present or absent. Following \cite{Padilla}, we also consider the ghost modes
for the radion and graviton in our theory.  In the case of negative bulk
gravitational constant, the range of parameters where the radion is ghost free
is rather narrow, and all massive gravitational modes have ghost-like nature.

We also calculate the effective gravitational potentials of static matter
sources on the visible and hidden brane in the generic case.

\section{The model}

The action of the model in the neighbourhood of one brane has the form
\begin{equation} \label{action}
S = M^3 \left[\int_{\rm bulk}\!\! \left( {\cal R} - 2 \Lambda \right) - 2
\int_{\rm brane}\!\! K \right] + \int_{\rm brane}\!\! \left( m^2 R - 2 \sigma
\right) + \int_{\rm brane}\!\! L \left( h_{ab}, \phi \right) \, ,
\end{equation}
where the first part, proportional to the cube of the bulk Planck mass $M^3$,
describes the bulk bounded by the brane, and the remaining integrals are taken
over the brane. Here, ${\cal R}$ is the curvature scalar in the bulk, $R$ is
the curvature scalar of the induced metric $h_{ab}$ on the brane, $K$ is the
trace of the tensor of the intrinsic curvature $K_{ab}$ of the brane with
respect to the inner normal, $L \left(h_{ab}, \phi \right)$ is the Lagrangian
of the matter fields $\phi$ on the brane, and integration in (\ref{action})
implies natural volume elements in the bulk and on the brane.  The action is
similar in the neighbourhood of the other brane.  In principle, two branes in
our model may have different Planck masses $m$, and, to allow for solutions
with flat vacuum branes, their tensions $\sigma$ must have opposite signs and
satisfy the well-known constraint \cite{RS1,RS2}
\begin{equation} \label{lambda-rs}
\Lambda_{\rm RS} \equiv {\Lambda \over 2} + {\sigma^2 \over 3 M^6} = 0 \, .
\end{equation}
Note that, in this paper, we allow for positive as well as negative signs of
the bulk Planck mass parameter $M$, the consequences of which will become
clear later.

Action (\ref{action}) leads to the bulk described by the usual Einstein
equation with cosmological constant:
\begin{equation} \label{bulk}
{\cal G}_{ab} + \Lambda g_{ab} = 0 \, ,
\end{equation}
while the field equation on the brane is
\begin{equation} \label{brane}
m^2 G_{ab} + \sigma h_{ab} = \tau_{ab} + M^3 \left(K_{ab} - h_{ab} K \right) \,
,
\end{equation}
where $\tau_{ab}$ is the stress--energy tensor on the brane stemming from
the last term in action (\ref{action}).

By contracting the Gauss identity
\begin{equation}\label{Gauss}
R_{abc}{}^d = h_a{}^f h_b{}^g h_c{}^k h^d{}_j {\cal R}_{fgk}{}^j + K_{ac}
K_b{}^d - K_{bc} K_a{}^d
\end{equation}
on the brane and using Eq.~(\ref{bulk}), one obtains the `constraint' equation
\begin{equation} \label{constraint}
R - 2 \Lambda + K_{ab} K^{ab} - K^2 = 0\, ,
\end{equation}
which, together with (\ref{brane}), implies the following closed scalar
equation on the brane:
\begin{equation} \label{closed}
M^6 \left( R - 2 \Lambda \right) + \left( m^2 G_{ab} + \sigma h_{ab} -
\tau_{ab} \right) \left( m^2 G^{ab} + \sigma h^{ab} - \tau^{ab} \right) - {1
\over 3} \left( m^2 R - 4 \sigma + \tau \right)^2 = 0\, ,
\end{equation}
where $\tau = h^{ab} \tau_{ab}$.

In the case of a vacuum brane ($\tau_{ab} = 0$), Eq.~(\ref{closed}) takes the
form
\begin{equation} \label{vacuum}
\left(M^6 + \frac23 \sigma m^2 \right) R + m^4 \left( R_{ab} R^{ab} - \frac13
R^2 \right) - 4 M^6 \Lambda_{\rm RS} = 0 \, ,
\end{equation}
where $\Lambda_{\rm RS}$ is given by Eq.~(\ref{lambda-rs}). It should be noted
that the second term in Eq.~(\ref{vacuum}) has {\em precisely\/} the form of
one of the terms in the expression for the conformal anomaly, which describes
the vacuum polarization at the one-loop level in curved space-time (see, e.g.,
\cite{BD}).\footnote{It is interesting that, while the conformal anomaly term
$R_{ab} R^{ab} - \frac13R^2$ cannot be obtained by the variation of a local
four-dimensional Lagrangian, the very same term is obtained via the variation
of a local Lagrangian in the five-dimensional braneworld theory under
investigation \cite{SS}.  Also note that this term is absent in the original
Randall--Sundrum model which has $m = 0$.}

Another useful relation is the Codazzi identity
\begin{equation}\label{Codazzi}
D_a \left(K^a{}_b - h^a{}_b K \right) = 0 \, ,
\end{equation}
which is valid at any timelike hypersurface in the bulk, in particular, on the
branes, due to Eq.~(\ref{bulk}).  Here, $D_a$ denotes the unique
covariant derivative on the timelike hypersurface associated with the induced
metric $h_{ab}$.

The gravitational equations in the bulk can be integrated by using
Gaussian normal coordinates, as described, e.g., in \cite{Shtanov}.
Specifically, in the Gaussian normal coordinates $(x, y)$, where $x =
\{x^\alpha\}$ are the coordinates on the brane and $y$ is the fifth coordinate
in the bulk, the metric is written as
\begin{equation}\label{bulk-metric}
d s^2 = dy^2 + h_{\alpha\beta} (x, y) dx^\alpha dx^\beta \, .
\end{equation}
Introducing also the tensor of extrinsic curvature $K_{ab}$ of every
hypersurface $y = {\rm const}$, one can obtain the following system
of differential equations for the components $h_{\alpha\beta}$ and
$K^\alpha{}_\beta$:
\begin{eqnarray} \label{s} {\partial K^\alpha{}_\beta \over \partial y}
&=& R^\alpha{}_\beta - K K^\alpha{}_\beta - \frac16 \delta^\alpha{}_\beta
\left( R + 2 \Lambda + K^\mu{}_\nu K^\nu{}_\mu - K^2 \right) \nonumber \\
&=& R^\alpha{}_\beta - K K^\alpha{}_\beta - \frac23
\delta^\alpha{}_\beta \Lambda \, ,
\end{eqnarray}
\begin{equation} \label{metric}
{\partial h_{\alpha\beta} \over \partial y} = 2 h_{\alpha\gamma}
K^\gamma{}_\beta \, ,
\end{equation}
where $R^\alpha{}_\beta$ are the components of the Ricci tensor of the metric
$h_{\alpha\beta}$ induced on the hypersurface $y = {\rm const}$, $R =
R^\alpha{}_\alpha$ is its scalar curvature, and $K = K^\alpha{}_\alpha$ is the
trace of the tensor of extrinsic curvature. The second equality in (\ref{s}) is
true by virtue of the `constraint' equation (\ref{constraint}).  Equations
(\ref{s}) and (\ref{metric}) together with the `constraint' equation
(\ref{constraint}) represent the $4\!+\!1$ splitting of the Einstein equations
in Gaussian normal coordinates. The initial conditions for these equations
are defined on the brane through Eq.~(\ref{brane}).

\section{Linear perturbations}

Linear perturbations of the RS1 model are well studied (see, e.g.,
\cite{linear,SV} and references therein).  Here, we would like to see the
modifications arising from the presence of the scalar-curvature terms in the
action for the branes (nonzero values of the masses $m$ and $m_*$). Several
aspects of this setup were also studied in \cite{CD,Padilla,Smolyakov,mirror}.
Our treatment in this and in the subsequent section is similar to that of
\cite{SV}.

The perturbed metric of our solution in Gaussian normal coordinates has the
form (\ref{bulk-metric}) with
\begin{equation} \label{induced}
h_{\alpha\beta} (x, y) =  a^2 (y) \Bigl[ \eta_{\alpha\beta} +
\gamma_{\alpha\beta} (x, y) \Bigr]\, ,
\end{equation}
where
\begin{equation} \label{scale}
\quad a(y) = \exp(-ky) \, , \quad k = {\sigma \over 3 M^3} \, ,
\end{equation}
and we emphasize that $k$ can be positive as well as negative depending on the
signs of $M$ and $\sigma$. The perturbations of the tensor of extrinsic
curvature and of the Einstein tensor have the  form
\begin{equation}\label{perturb}
\delta K^\alpha{}_\beta = \frac12 {\partial \gamma^\alpha{}_\beta \over
\partial y} \, , \quad  G_{\alpha\beta} = - \frac12 \Box \bar
\gamma_{\alpha\beta} + \partial^\gamma \partial_{(\alpha} \bar
\gamma_{\beta)\gamma} - \frac12 \eta_{\alpha\beta} \partial^\gamma
\partial^\delta \bar \gamma_{\gamma\delta}\, ,
\end{equation}
where $\bar \gamma_{\alpha\beta} = \gamma_{\alpha\beta} - \frac12
\eta_{\alpha\beta} \gamma$, $\gamma = \gamma^\alpha{}_\alpha$, and $\Box =
\partial^\alpha\partial_\alpha$.  Here and below, the indices of
$\partial_\alpha$ and $\gamma_{\alpha\beta}$ are raised and lowered with
respect to the flat metric $\eta_{\alpha\beta}$.

Using the freedom of choice of the coordinates $x^\alpha$ on the brane, one can
choose the harmonic gauge in which $\partial^\alpha \bar \gamma_{\alpha\beta} =
0$ on one of the branes.  In this gauge, we have
\begin{equation}\label{harmoncor}
G_{\alpha\beta} = - \frac12 \Box \bar \gamma_{\alpha\beta}
\end{equation}
on that brane.

In the unperturbed solution, the first (visible) brane is assumed to be at $y =
0$, and the second (hidden) brane is at $y = \rho$.  First, we consider the
situation where the hidden brane does not have matter on it (stress--energy
tensor equal to zero).  Then, when studying perturbations, it is convenient to
choose Gaussian normal coordinates with respect to the hidden brane. Thus, the
hidden brane remains at $y = \rho$, while the position of the visible brane is
linearly perturbed to become $y = \phi(x)$, which is the so-called radion
degree of freedom.  Let $m$ and $\sigma$ denote the Planck mass and tension of
the visible brane, and let those of the hidden brane be $m_*$ and $\sigma_* = -
\sigma$, respectively. The linearly perturbed boundary equation (\ref{brane})
on the second (hidden) brane becomes
\begin{equation}
- m_*^2 G^\alpha{}_\beta = M^3 \delta S^\alpha{}_\beta \, ,
\end{equation}
where $S^\alpha{}_\beta = K^\alpha{}_\beta - \delta^\alpha{}_\beta K$, and
we have taken into account that the extrinsic curvature is calculated
with respect to the normal in the positive direction of $y$.  Choosing harmonic
coordinates on the hidden brane, for which (\ref{harmoncor}) is satisfied,
we have
\begin{equation}\label{hidden}
{m_*^2 \over a_*^2} \Box \bar \gamma^\alpha{}_\beta = M^3 \left( {\partial \bar
\gamma^\alpha{}_\beta \over \partial y} + \frac12 \delta^\alpha{}_\beta
{\partial \bar \gamma \over \partial y} \right) \, ,
\end{equation}
where $a_* = a(\rho) = e^{-k\rho}$ and $\bar \gamma = \bar
\gamma^\alpha{}_\alpha$.

Linearization of the vacuum constraint equation (\ref{vacuum}) implies the
condition $\Box \bar \gamma = 0$ on the hidden brane if
\begin{equation} \label{nonsin}
M^6 + \frac23 \sigma_* m_*^2 \ne 0 \, ,
\end{equation}
which we assume to be the case.  This condition and Eq.~(\ref{hidden}) implies
the condition $\partial \bar \gamma /
\partial y = 0$ at the hidden brane.  Then the Codazzi relation
(\ref{Codazzi}) implies the condition $\partial \left(\partial^\alpha \bar
\gamma_{\alpha\beta}\right) / \partial y = 0$ at the same brane.

Now we turn to Eqs.~(\ref{s}) and (\ref{metric}).  Using (\ref{perturb}), we
can write the second-order differential equations for perturbations $\bar
\gamma_{\alpha\beta}$ in the bulk. First, we verify that the Gaussian normal
coordinates $x^\alpha$ remain harmonic in the bulk. We introduce the quantity
\begin{equation}
v^\alpha = \partial_\beta \bar \gamma^{\beta\alpha} \, ,
\end{equation}
which is an indicator of the harmonicity of the coordinates $x^\alpha$ on the
hypersurface $y = {\rm const}$. Then we can write the system of differential
equations for $v^\alpha$ and $\bar \gamma$ that stems from system (\ref{s}),
(\ref{metric}):
\begin{equation}\label{harmon}
{\partial^2 v^\alpha \over \partial y^2} = 4k {\partial v^\alpha \over \partial
y} + k {\partial \left(\partial^\alpha \bar \gamma \right) \over \partial y} \,
, \qquad {\partial^2 \bar \gamma \over \partial y^2} = - \frac{1}{a^2} \left( 2
\partial_\alpha v^\alpha + \Box \bar \gamma \right) + 8k {\partial
\bar \gamma \over \partial y} \, ,
\end{equation}
with the following boundary conditions at the hidden brane ($y = \rho$):
\begin{equation}\label{init-harmon}
v^\alpha = 0 \, , \quad {\partial v^\alpha \over \partial y} = 0 \, , \quad
\Box \bar \gamma = 0 \, , \quad {\partial \bar \gamma \over \partial y} = 0 \,
.
\end{equation}
The unique solution of system (\ref{harmon}) in the bulk with the boundary
conditions (\ref{init-harmon}) is
\begin{equation}\label{harmonrel}
v^\alpha (x, y) \equiv 0 \, , \quad \bar \gamma (x, y) \equiv \bar \gamma (x)
\, , \quad \Box \bar \gamma (x) = 0 \, .
\end{equation}
In particular, this means that the Gaussian normal coordinates which we are using
remain harmonic with respect to $x$ all over the bulk.

Taking into account relations (\ref{harmonrel}), from (\ref{s}), (\ref{metric})
one obtains the system of equations for perturbations in the bulk:
\begin{equation}\label{eq-bulk}
{\partial^2 \bar \gamma_{\alpha\beta} \over \partial y^2} - 4k {\partial \bar
\gamma_{\alpha\beta} \over \partial y} + {1 \over a^2} \Box \bar
\gamma_{\alpha\beta} = 0
\end{equation}
with the boundary condition at $y = \rho$ which stems from (\ref{hidden}):
\begin{equation}\label{b-hidden}
{\partial \bar \gamma_{\alpha\beta} \over \partial y} = {m_*^2 \over M^3 a_*^2}
\Box \bar \gamma_{\alpha\beta} \, .
\end{equation}

To obtain the boundary equations on the visible brane, one must take into
account its `bending' in the bulk: $y = \phi(x)$.  The induced metric on the
visible brane in the linear approximation becomes
\begin{equation}\label{indmet}
h^{\rm vis}_{\alpha\beta} = (1 - 2 k \phi) \eta_{\alpha\beta} +
\gamma_{\alpha\beta} \, ,
\end{equation}
so that its perturbation is
\begin{equation}
\quad \gamma^{\rm vis}_{\alpha\beta} = \gamma_{\alpha\beta} - 2 k \phi
\eta_{\alpha\beta} \, , \quad \bar \gamma^{\rm vis}_{\alpha\beta} = \bar
\gamma_{\alpha\beta} + 2 k \phi \eta_{\alpha\beta} \, .
\end{equation}
Substituting it to the boundary condition at the visible brane
\begin{equation}
m^2 G^\alpha{}_\beta = M^3 \delta S^\alpha{}_\beta + \tau^\alpha{}_\beta \, ,
\end{equation}
we obtain the boundary condition at $y = 0$:
\begin{equation}\label{g}
m^2 G_{\alpha\beta} \equiv -{m^2 \over 2} \Box \bar \gamma_{\alpha\beta} + 2
m^2 k \left( \partial_\alpha \partial_\beta - \eta_{\alpha\beta} \Box \right)
\phi = \tau_{\alpha\beta} + M^3 \left( \eta_{\alpha\beta} \Box -
\partial_\alpha \partial_\beta \right) \phi + \frac12 M^3 {\partial \bar
\gamma_{\alpha\beta} \over \partial y} \, .
\end{equation}
Taking trace of this equation, we obtain the equation for the radion field $\phi$:
\begin{equation} \label{radion}
- 3 A \Box \phi = \tau \, ,
\end{equation}
where $\tau \equiv \eta^{\alpha\beta} \tau_{\alpha\beta}$ is the trace of the
stress--energy tensor, and $A = M^3 + 2 k m^2$.  Thus, the radion field is coupled
to the trace of the stress--energy tensor, as is the case in the Randall--Sundrum
model \cite{linear}, but with different coupling constant.  Using Eq.~(\ref{radion}),
from (\ref{g}) we obtain
\begin{equation}\label{b-visible}
-{m^2 \over 2} \Box \bar \gamma_{\alpha\beta} = \tau_{\alpha\beta} - \frac13
\eta_{\alpha\beta} \tau - A \partial_\alpha \partial_\beta \phi + \frac12 M^3
{\partial \bar \gamma_{\alpha\beta} \over
\partial y} \, .
\end{equation}

Now we have to solve the bulk equations (\ref{eq-bulk}) with the boundary
conditions (\ref{radion}), (\ref{b-hidden}) and (\ref{b-visible}). Proceeding
to the Fourier transform with momenta $p_\alpha$ in the coordinates $x^\alpha$
and omitting the tensor indices, we have for the Fourier image $\psi (q, y)$ of
$\bar \gamma_{\alpha\beta} (x, y)$:
\begin{equation}\label{eq-psi}
\psi'' - 4 k \psi' + q^2 e^{2 k y} \psi = 0 \, ,
\end{equation}
where the prime denotes the derivative with respect to $y$, and $q = \sqrt{-
p^2}$ (here we assume $p^2 \equiv p^\alpha p_\alpha \le 0$; the tachyonic case
will be studied in Sec.~\ref{tachyon}). After the standard change of variable
and function
\begin{equation}
z (y) = {q  e^{ky} \over k} \, , \qquad \psi (z) = z^2 \chi (z) \, ,
\end{equation}
we get the equation
\begin{equation}\label{eq-chi}
z^2 \chi'' + z \chi' + \left(z^2 - 4\right) \chi = 0 \, ,
\end{equation}
in which the prime denotes the derivative with respect to $z$.  Note that $z
(y)$ is a monotonic function of $y$ for both signs of $k$, but the sign of $z$
coincides with the sign of the constant $k$.  The boundary conditions follow
from (\ref{b-visible}) and (\ref{b-hidden}):
\begin{eqnarray}\label{b1-vis}
- m^2 k^2 z_0^2 \chi (z_0) &=& {2 T \over z_0^2} + k M^3 \Bigl[ z_0 \chi' (z_0)
+ 2 \chi (z_0) \Bigr] \, , \quad z_0 = z(0) = \frac{q}{k} \, , \\
\label{b1-hid} m_*^2  k^2 z_*^2 \chi (z_*) &=& k M^3 \Bigl[ z_* \chi' (z_*) + 2
\chi (z_*) \Bigr] \, , \quad z_* = z(\rho) = \frac{q}{k} e^{k\rho} \, ,
\end{eqnarray}
where $T$ stands for the Fourier transform of the expression
\begin{equation} \label{t}
T_{\alpha\beta} = \tau_{\alpha\beta} - \frac13 \eta_{\alpha\beta} \tau -
A \partial_\alpha \partial_\beta \phi
\end{equation}
with tensor indices omitted.

The general solution of the Bessel equation (\ref{eq-chi}) is given by
\begin{equation}\label{chi}
\chi (z) = P J_2 \Bigl(|z|\Bigr) + Q Y_2 \Bigl(|z|\Bigr) \, ,
\end{equation}
where $J_2$ and $Y_2$ are the Bessel functions, $P$ and $Q$ are constants, and
the modulus of $z$ reflects the fact that the domain of $z$ is positive or
negative depending on the sign of $k$.

Using the recurrence relations
\begin{equation}\label{recurrence}
z J'_2 (z) + 2 J_2 (z) = z J_1 (z) \, , \quad z Y'_2 (z) + 2 Y_2 (z) = z Y_1
(z) \, ,
\end{equation}
we obtain from (\ref{b1-vis}) and (\ref{b1-hid}):
\begin{eqnarray}\label{b2-vis}
- m^2 k^2 z_0^2 \left[ P J_2 \Bigl(|z_0|\Bigr) + Q Y_2 \Bigl(|z_0|\Bigr)
\right] &=& {2 T \over z_0^2} + k M^3 |z_0| \left[ P J_1 \Bigl(|z_0|\Bigr) + Q
Y_1 \Bigl(|z_0|\Bigr) \right] \, , \\ \label{b2-hid}
m_*^2  k^2 z_*^2 \left[ P J_2 \Bigl(|z_*|\Bigr) + Q Y_2 \Bigl(|z_*|\Bigr)
\right] &=& k M^3 |z_*| \left[ P J_1 \Bigl(|z_*|\Bigr) + Q Y_1 \Bigl(|z_*|\Bigr)
\right] \, ,
\end{eqnarray}
solving which, one finds the constants $P$ and $Q$ and obtains the solution for
$\psi (z)$:
\begin{equation}\label{psi}
\psi (z) = \left({6 T z^2 \over \sigma |z_0|^3}\right) {C_Y^* J_2 \Bigl( |z|
\Bigr) - C_J^* Y_2 \Bigl( |z| \Bigr) \over C_Y^0 C_J^* - C_J^0 C_Y^*} \, ,
\end{equation}
where the constants are given by
\begin{equation}\label{cs} \begin{array}{l}
C_Y^0 = Y_1 \Bigl( |z_0| \Bigr) + \displaystyle {m^2 \over M^3} k |z_0| Y_2
\Bigl( |z_0| \Bigr) \, , \quad C_J^0 = J_1 \Bigl( |z_0| \Bigr) + {m^2 \over
M^3} k |z_0| J_2 \Bigl( |z_0| \Bigr) \, ,  \medskip \\ C_Y^* = Y_1 \Bigl( |z_*|
\Bigr) - \displaystyle  {m_*^2 \over M^3} k |z_*| Y_2 \Bigl( |z_*| \Bigr) \, ,
\quad C_J^* = J_1 \Bigl( |z_*| \Bigr) - {m_*^2 \over M^3} k |z_*| J_2 \Bigl(
|z_*| \Bigr) \, .
\end{array}\end{equation}
These results differ from the similar results \cite{linear,SV} of the RS1 model
by the presence of the terms containing the brane Planck masses $m$ and $m_*$
in Eqs.~(\ref{cs}).

The spectrum of the model is determined by the equality of the denominator of
(\ref{psi}) to zero.  Introducing the dimensionless variable $s = q/|k|$ and
parameters $\mu = k m^2 / M^3$, $\mu_* = k m_*^2 / M^3$, and $\alpha =
e^{k\rho}$, we obtain the following equation for the spectrum:
\begin{equation}\label{spec}
F_1(s) + \mu s F(s) + \alpha \mu_* s F_*(s) + \alpha \mu \mu_* s^2
F_2 (s) = 0 \, ,
\end{equation}
where
\begin{equation}\begin{array}{l}\label{fs}
F_1(s) = J_1 (s) Y_1 (\alpha s) - J_1 (\alpha s) Y_1 (s)  \, , \\ F(s) = J_2
(s) Y_1 (\alpha s) - J_1 (\alpha s) Y_2 (s)  \, , \\ F_*(s) = J_2 (\alpha s)
Y_1 (s) - J_1 (s) Y_2 (\alpha s) \, , \\ F_2(s) = J_2 (\alpha s) Y_2 (s) - J_2
(s) Y_2 (\alpha s)  \, .
\end{array}\end{equation}

If both masses $m$ and $m_*$ are nonzero, then the ultraviolet asymptotics of
the spectrum for the Kaluza--Klein modes is determined by the zeros of the
last term in (\ref{spec}), so that
\begin{equation}
s_n \sim {\pi n \over \alpha - 1} \, , \quad n \gg 1 \, ,
\end{equation}
which coincides with the asymptotics of the spectrum in the Randall--Sundrum
model, determined by the zeros of the first term in (\ref{spec}).

If $m \ne 0$, $m_* = 0$, then the asymptotics of the spectrum is determined by
the second term in (\ref{spec}):
\begin{equation}
s_n \sim {\pi n - \frac{\pi}{2}\over \alpha - 1} \, , \quad n \gg 1 \, .
\end{equation}
If $m = 0$, $m_* \ne 0$, then it is determined by the third term in (\ref{spec}):
\begin{equation}
s_n \sim {\pi n + \frac{\pi}{2}\over \alpha - 1} \, , \quad n \gg 1 \, .
\end{equation}
In these last two cases, the spectrum is somewhat shifted.

\section{Linearized equations in the zero-mode approximation}

In the zero-mode approximation \cite{linear,SV}, one considers the limit
as $q \to 0$.  In this limit, using Eqs.~(\ref{g}), (\ref{radion}), and (\ref{psi})
and expanding all functions of $q = \sqrt{- p^2}$ in powers of $q$, we obtain
the linearized gravity equation in the case of matter present only on
the visible brane:
\begin{equation}\label{g-vis}
G_{\alpha\beta} = {2 k \over A - B e^{- 2 k\rho}} \left[ \tau_{\alpha\beta} -
{B e^{-2 k\rho} \over 3 A} \left( \eta_{\alpha\beta} - {p_\alpha p_\beta \over
p^2} \right) \tau \right] + {\cal O} \left(p^2\right) \, ,
\end{equation}
where it should be stressed that $G_{\alpha\beta}$  is the Einstein tensor of the
induced metric (\ref{indmet}) on the brane, and the constants $A$ and $B$ are
given by\footnote{The constant $A$ is the same as in Eq.~(\ref{radion}).}
\begin{equation}
\label{AB}
A = M^3 + 2 k m^2 \, , \quad B = M^3 - 2 k m_*^2 \, .
\end{equation}

The effective Newton's constant $G_{\rm N}$ is given by the relation
\begin{equation} \label{newton}
8 \pi G_{\rm N} = {2 k \over A - B e^{- 2 k\rho}} \, ,
\end{equation}
and one should note the extra contribution from the radion in (\ref{g-vis}),
which involves  the trace of the stress--energy tensor.\footnote{In
Eq.~(\ref{g-vis}) as well as in similar equations of this section, we formally
express the radion field through the trace of the stress--energy tensor using
Eq.~(\ref{radion}), similarly to how it is done, e.g., in \cite{SV}.} If $k >
0$, this contribution is exponentially suppressed for large separations between
the branes, $k\rho \gg 1$.

If matter is present only on the hidden brane, then it still induces
curvature on the visible brane \cite{linear,SV}.  In our theory, we obtain the result
\begin{equation}\label{g-hid}
G_{\alpha\beta} = {2 k \over A e^{2 k\rho} - B } \left[ \tau^*_{\alpha\beta} -
{e^{- 2 k\rho} \over 3} \left( \eta_{\alpha\beta} - {p_\alpha p_\beta \over
p^2} \right) \tau^* \right] + {\cal O} \left(p^2\right) \, ,
\end{equation}
where $\tau^*_{\alpha\beta}$ is the stress--energy tensor on the hidden brane,
and $\tau^*$ is its trace.
If both branes contain matter, then the results on the right-hand sides
of (\ref{g-vis}) and (\ref{g-hid}) simply add together.

A few comments are in order about the obtained results.  First of all, in the
limit of zero Planck masses for the branes, $m = m_* = 0$, we have $A = B =
M^3$, and they turn to the results previously obtained for the Randall--Sundrum
two-brane model \cite{linear,SV}.  The presence of two new mass parameters $m$
and $m_*$ extends the freedom of the model.  Thus, if the constant $B$ turns
out to be sufficiently small, then the scalar contribution to the right-hand
side of (\ref{g-vis}), proportional to the trace of the stress--energy tensor,
may become negligibly small.  Note, however, that it is not possible to set
either the constant $A$ or the constant $B$ exactly to zero in our expressions
since, in this case, the theory becomes singular.  This can be seen, e.g., from
Eq.~(\ref{vacuum}), in which the first (linear in curvature) term is
proportional to $A$ [the same property is observed in the general equation
(\ref{closed})].  In particular, the nonzero value of the constant $B$ was
already assumed in the linearization scheme [see Eq.~(\ref{nonsin})]. The
special cases where either $A$ or $B$ is equal to zero must be studied
separately. Some results in this direction were recently reported in
\cite{Smolyakov}, where it was pointed out that the linearized theory possesses
some additional symmetry in this case.

To see how this degeneracy arises in some more detail, we turn to the Gauss
identity (\ref{Gauss}) again and, following the procedure first employed in
\cite{SMS}, contract it once on the brane using equations (\ref{bulk}) and
(\ref{brane}). We obtain the effective equation on the brane that generalizes
the result of \cite{SMS} to the presence of the brane curvature term:
\begin{equation}\label{effective}
G_{ab} + \Lambda_{\rm RS} {M^3 \over A} h_{ab} = {2 \sigma \over 3 M^3 A}
\tau_{ab} + {1 \over M^3 A} Q_{ab} - {M^3 \over A} W_{ab} \, ,
\end{equation}
where $\Lambda_{\rm RS}$ is given by (\ref{lambda-rs}),
\begin{equation}
Q_{ab} = \frac13 E E_{ab} - E_{ac} E^{c}{}_b + \frac12 \left(E_{cd} E^{cd} -
\frac13 E^2 \right) h_{ab}
\end{equation}
is the quadratic expression with respect to the `bare' Einstein equation
$E_{ab} \equiv m^2 G_{ab} - \tau_{ab}$ on the brane, $E = h^{ab} E_{ab}$, and
$W_{ab} \equiv h^c{}_a h^e{}_b W_{cdef} n^d n^f$ is a projection of the bulk
Weyl tensor $W_{abcd}$ to the brane. One can see that {\em all\/} the couplings
in (\ref{effective}), including the effective cosmological and gravitational
constants, are inversely proportional to the constant $A$, which indicates that
the theory becomes degenerate in the case $A = 0$.  In the absence of the
curvature term on the brane ($m = 0$), we have $A = M^3$, which brings us to
the original result of \cite{SMS}.

Our second remark is that, unlike in the original Randall--Sundrum model ($m =
m_* = 0$), in our theory the sign of the constant $M$ is not fixed by the
zero-mode approximation: apart from the scalar contribution described by the
trace of the stress--energy tensor, matching with the general-relativity limit
fixes only the sign of the overall constant in (\ref{g-vis}) and (\ref{g-hid}).
In particular, for a sufficiently small absolute value of $M$, namely, $|M^3 /
2 k| \ll m^2,\, m_*^2\,$, the sign of $M$ does not matter. In the formal limit
$M \to 0$ with $k > 0$ (hence, $k \to + \infty$), the equation for the visible
brane (\ref{g-vis}) turns to the usual linearized Einstein equation. In the
formal simultaneous limit $M \to 0$ and $\sigma \to 0$ so that $k = \sigma /
M^3$ is fixed, expressions (\ref{g-vis}) and (\ref{g-hid}) become
\begin{equation}\label{lim-vis}
G_{\alpha\beta} = {1 \over m^2 + m_*^2 e^{- 2 k\rho}} \left[ \tau_{\alpha\beta}
+ {m_*^2 e^{-2 k\rho} \over 3 m^2} \left( \eta_{\alpha\beta} - {p_\alpha
p_\beta \over p^2} \right) \tau \right] + {\cal O} \left(p^2\right)
\end{equation}
and
\begin{equation}\label{lim-hid}
G_{\alpha\beta} = {1 \over m^2 e^{2 k\rho} + m_*^2 } \left[
\tau^*_{\alpha\beta} - {e^{- 2 k\rho} \over 3} \left( \eta_{\alpha\beta} -
{p_\alpha p_\beta \over p^2} \right) \tau^* \right] + {\cal O} \left(p^2\right)
\, ,
\end{equation}
respectively.  However, in the following section we will see that the massive
gravitational modes in the theory with negative value of $M$ have ghost-like
nature.

Finally, we note that our result does not explicitly contain the constant
$\sigma$ but contains it only in the combination $k = \sigma / 3 M^3$.
Therefore, for one and the same effective law of gravity (\ref{g-vis}) and
(\ref{g-hid}), the visible brane can have either positive or negative brane
tension, depending on the sign of $M$.  In particular, the zero-mode graviton
is `localized' around the visible brane ($k > 0$) even if its tension is
negative, provided $M$ is also negative. If $k > 0$, then, in the limit $\rho
\to \infty$, we pass to the one-brane model in Eq.~(\ref{g-vis}), which has the
form of the corresponding equation of general relativity.

\section{Tachyonic modes and ghosts}
 \label{tachyon}

In the original Randall--Sundrum model, negative values of the bulk Planck mass
$M$ are nonphysical because this leads to negative effective Newton's constant.
This can be seen by setting $m = m_* = 0$ in Eq.~(\ref{newton}), thus having $A
= B = M^3$ in it.  If $M < 0$, then the effective Newton's constant is negative
for any sign of $k$, which means that the massless graviton becomes a ghost.

The presence of the curvature terms in the action for the brane relaxes the
situation with the massless gravitational modes and thus relaxes the necessity
of dealing only with positive values of $M$. Negative values of $M$ are of
interest in view of some of the braneworld cosmological models with {\em
negative\/} brane tension, in particular, the model of disappearing dark energy
\cite{SS,AS}, as discussed in the introduction. However, unlike in the pure
Randall--Sundrum case, the presence of the curvature terms in the action for
the branes leads to a possibility of unwanted tachyonic modes  and ghost-like
character of the massive modes in the gravitational sector of the theory if $M
< 0$. In this section, we demonstrate that there can be only one or two
tachyonic mass eigenvalues in the theory under consideration and determine the
range of parameters for which tachyonic modes do not exist.  We also show that
the massive gravitational modes have ghost-like character in the case of
negative $M$.

In looking for tachyonic modes, one needs to solve Eq.~(\ref{eq-psi}) for $q^2
= - p^2 < 0$, i.e.,
\begin{equation}\label{eqt-psi}
\psi'' - 4 k \psi' - p^2 e^{2 k y} \psi = 0 \, , \quad p^2 > 0 \, ,
\end{equation}
where the prime denotes the derivative with respect to $y$. After the standard
change of variable and function
\begin{equation}
z (y) = p e^{ky} / k \, , \qquad \psi (z) = z^2 \chi (z) \, ,
\end{equation}
we get the equation for the new function $\chi(z)$:
\begin{equation}\label{eqt-chi}
z^2 \chi'' + z \chi' - \left(z^2 + 4\right) \chi = 0 \, ,
\end{equation}
in which the prime denotes the derivative with respect to $z$.  With matter
present on the visible brane only, the boundary conditions are similar to
(\ref{b1-vis}) and (\ref{b1-hid}):
\begin{eqnarray}\label{bt-vis}
m^2 k^2 z_0^2 \chi (z_0) &=& {2 T \over z_0^2} + k M^3 \Bigl[ z_0 \chi' (z_0) +
2 \chi (z_0) \Bigr] \, , \quad z_0 = z(0) = \frac{p}{k} \, ,
\\ \label{bt-hid}
- m_*^2  k^2 z_*^2 \chi (z_*) &=& k M^3 \Bigl[ z_* \chi' (z_*) +  2 \chi (z_*)
\Bigr] \, , \quad z_* = z(\rho) = \frac{p}{k} e^{k\rho} \, ,
\end{eqnarray}
where $T$ is the Fourier transform of expression (\ref{t}), with tensor indices
omitted.

Solution of (\ref{eqt-chi}) is now given by the modified Bessel functions $I_2$
and $K_2$:
\begin{equation}\label{t-chi}
\chi (z) = P I_2 \Bigl(|z|\Bigr) + Q K_2 \Bigl(|z|\Bigr) \, ,
\end{equation}
where $P$ and $Q$ are constants, and the modulus of $z$ again reflects the fact
that the domain of $z$ is positive or negative depending on the sign of $k$.
The recurrence relations of type (\ref{recurrence}) are valid also for the
modified Bessel functions:
\begin{equation}\label{recur-mod}
z I'_2 (z) + 2 I_2 (z) = z I_1 (z) \, , \quad z K'_2 (z) + 2 K_2 (z) = - z K_1
(z) \, ,
\end{equation}
and we can use them in deriving the solution similar to (\ref{psi}):
\begin{equation}\label{psi-t}
\psi (z) = \left({6 T z^2 \over \sigma |z_0|^3}\right) {C_K^* I_2 \Bigl( |z|
\Bigr) + C_I^* K_2 \Bigl( |z| \Bigr) \over C_I^* C_K^0 - C_I^0 C_K^*} \, ,
\end{equation}
where
\begin{equation}\label{Cs}\begin{array}{l}
C_I^0 = I_1 \Bigl( |z_0| \Bigr) - \displaystyle {m^2 \over M^3} k |z_0| I_2
\Bigl( |z_0| \Bigr) \, , \quad C_K^0 = K_1 \Bigl( |z_0| \Bigr) +  {m^2 \over
M^3} k |z_0| K_2 \Bigl( |z_0| \Bigr)  \medskip \\ C_I^* = I_1 \Bigl( |z_*|
\Bigr) + \displaystyle {m_*^2 \over M^3} k |z_*| I_2 \Bigl( |z_*| \Bigr) \, ,
\quad C_K^* = K_1 \Bigl( |z_*| \Bigr) - {m_*^2 \over M^3} k |z_*| K_2 \Bigl(
|z_*| \Bigr) \, .
\end{array}\end{equation}

At this point, we note that the restriction to the brane at $y = 0$ and the
limit of brane separation $\rho \to \infty$ brings expression (\ref{psi-t}) to
the form obtained for the one-brane case in \cite{Tanaka}. Our result
generalizes it to the case of two branes with arbitrary sign of the bulk
gravitational constant.

Tachyonic modes correspond to those values of $p$ for which the denominator of
(\ref{psi-t}) turns to zero:
\begin{equation}
C_I^* C_K^0 - C_I^0 C_K^* = 0 \, .
\end{equation}
Since the transformation  $p \to e^{- k\rho}\!p$ followed by $k \rightarrow -
k$ and $m \leftrightarrow m_*$ does not change the spectrum of the theory, it
is sufficient to study only the case $k > 0$.  Using the dimensionless variable
$s = p/k$ and parameters $\mu = k m^2 /M^3$, $\mu_* = k m_*^2 /M^3$, and
$\alpha = e^{k\rho}$,  we obtain the equation for tachyonic modes:
\begin{equation}\label{E}
E(s) \equiv D_1(s) + \mu s D(s) + \alpha \mu_* s D_*(s) + \alpha \mu \mu_* s^2
D_2 (s) = 0 \, ,
\end{equation}
where
\begin{equation}\begin{array}{l}\label{ds}
D_1(s) = I_1 (\alpha s) K_1 (s) - I_1 (s) K_1 (\alpha s)  \, ,  \\ D(s) = I_2
(s) K_1 (\alpha s) + I_1 (\alpha s) K_2 (s)  \, ,  \\ D_*(s) = I_1 (s) K_2
(\alpha s) + I_2 (\alpha s) K_1 (s) \, , \\ D_2(s) = I_2 (\alpha s) K_2 (s) -
I_2 (s) K_2 (\alpha s)  \, .
\end{array}\end{equation}

Since $\alpha > 1$ for $k > 0$, all functions in (\ref{ds}) are strictly
positive for positive $s$.  This implies that tachyonic modes are absent in the
case $\mu,\,\mu_* > 0$, or, equivalently, $M > 0$.

However, tachyonic modes may be present in the opposite case $M <
0$.  It is possible to indicate the corresponding range of parameters where
tachyonic modes are present or absent.  To do this, it is convenient to introduce
the following auxiliary function of two variables $s$ and $\bar s$:
\begin{equation}\label{E2}
\bar E(s, \bar s) \equiv D_1(s) + \mu \bar s D(s) + \alpha \mu_* \bar s D_*(s)
+ \alpha \mu \mu_*  \bar s^2
 D_2 (s)
\end{equation}
[to be compared with (\ref{E})].  By construction, $\bar E (s, s) \equiv E(s)$.

First, we consider the case where both $\mu$ and $\mu_*$ are nonzero (in the
present case, they are then both negative). Then the equation
\begin{equation}\label{bareq}
\bar E (s, \bar s) = 0
\end{equation}
gives the two branches of solutions with respect to $\bar s$:
\begin{equation}\label{bars}
\bar s_\pm (s) = {\Bigl|\mu D(s) + \alpha \mu_* D_*(s) \Bigr| \pm
\sqrt{\Bigl(\mu D(s) + \alpha \mu_* D_*(s)\Bigr)^2 - 4 \alpha \mu \mu_*
D_1(s) D_2(s) } \over 2 \alpha \mu \mu_* D_2 (s) } \, ,
\end{equation}
and solving the original equation (\ref{E}) is equivalent to solving
one of the equations
\begin{equation}\label{eqs}
\bar s_\pm (s) = s \, .
\end{equation}

It is easy to verify that the expression under the square root of (\ref{bars})
is strictly positive for positive $s$ so that the two solutions $\bar s_\pm
(s)$ exist for all $s >0$ and are positive.  The asymptotic behavior of these
solutions for small and large $s$ can easily be found:
\begin{equation}
\bar s_+ (s) \sim {4 \Bigl| \alpha^2 \mu  + \mu_* \Bigr| \over
\left(\alpha^4 - 1\right) \mu \mu_* s } \, , \quad
\bar s_-(s) \sim {\left(\alpha^2 - 1\right) s \over 2 \Bigl| \alpha^2 \mu
+ \mu_* \Bigr| } \, , \qquad s \to 0 \, ,
\end{equation}
\begin{equation}
\bar s_\pm (s) \to {\Bigl| \mu + \alpha \mu_* \Bigr| \pm \Bigl| \mu -
\alpha \mu_* \Bigr| \over 2 \alpha \mu \mu_*} = {\rm const} \, , \qquad
s \to \infty \, .
\end{equation}

From these expressions it is clear that the graph of $\bar s_+(s)$ crosses
the graph of $f(s) = s$ at least once for any values of parameters in the range
$\mu, \mu_* < 0$ under consideration.  Thus, Eq.~(\ref{eqs}) has a solution,
and at least one tachyonic mass eigenvalue is present in this range of
parameters.  Numerical computation
indicates that there is exactly one solution connected with the branch
$\bar s_+(s)$ in this case.

It is also clear that the graph of $\bar s_-(s)$ definitely crosses
the graph of $f(s) = s$ in the case where $\bar s'_-(0) > 1$, or
\begin{equation}\label{s_}
{\left(\alpha^2 - 1\right) \over 2 \Bigl| \alpha^2  \mu
+ \mu_* \Bigr| } > 1 \, .
\end{equation}
Again, numerical computation indicates that there is only one tachyonic
solution connected with the branch $\bar s_-(s)$ in this case.  They also
indicate that tachyonic modes connected with the branch $\bar s_-(s)$ are
absent in the case of the opposite inequality in (\ref{s_}).

Thus, tachyonic modes exist for all values of parameters in the range
$\mu, \mu_* < 0$.  In the case under consideration, $M < 0$, one can
expect tachyonic modes to be absent only if one of
the brane Planck masses $m$ or $m_*$ is zero.\footnote{Tachyonic modes are
obviously absent in the Randall--Sundrum model ($m = m_* = 0$) even in
the case of negative bulk gravitational constant ($M < 0$), but this model
is already excluded as resulting in negative effective Newton's constant
on the brane (see the beginning of this section).}   We show that this is
indeed the case and determine the range of masses and brane separations
for which tachyonic modes are absent.

In the case $\mu < 0$, $\mu_* = 0$, the function $\bar E(s, \bar s)$
given by Eq.~(\ref{E2}) takes the simple form
\begin{equation}
\bar E(s, \bar s) \equiv D_1(s) + \mu \bar s D(s) \, ,
\end{equation}
and Eq.~(\ref{bareq}) has a single solution with respect to $\bar s$:
\begin{equation}
\bar s (s) = - {D_1(s) \over \mu D(s)} = {D_1 (s) \over |\mu| D(s)} \, .
\end{equation}

It can be verified that the function $D_1(s)/D(s)$ is
convex upwards, so that the equation $\bar s (s) = s$ has exactly one
solution or no solutions in the range $s > 0$ depending on the value of
the derivative $\bar s'(0)$.  Specifically, a solution exists if
$\bar s'(0) > 1$, and there are no solutions in the opposite case
$\bar s'(0) \le 1$.  Calculating the derivative $\bar s'(0)$, we obtain
that exactly one tachyonic mass eigenvalue is present in the theory if
\begin{equation}
|\mu| < \frac12 \left(1 - \alpha^{-2} \right) \, ,
\end{equation}
and tachyonic modes are absent if the value of the Planck mass $m$ is
sufficiently large, namely, if
\begin{equation}\label{mass}
|\mu| \ge \frac12 \left(1 - \alpha^{-2} \right) \, .
\end{equation}
In the limit of infinite separation between branes, $\alpha = e^{k\rho} \to
\infty$, the condition of absence of tachyonic modes becomes $|\mu| \ge 1/2$,
which coincides with the condition $A \ge 0$, where $A$ is given by (\ref{AB}).
Interestingly, this is also the condition of positivity of the effective
Newton's constant in the zero-mode approximation (\ref{g-vis}), (\ref{g-hid})
in the same limit.

The case $\mu = 0$, $\mu_* < 0$ is analyzed in quite a similar way. Now the
function $\bar E(s, \bar s)$ given by Eq.~(\ref{E2}) takes the form
\begin{equation}
\bar E(s, \bar s) \equiv D_1(s) + \alpha \mu_* \bar s D_*(s) \, ,
\end{equation}
and Eq.~(\ref{bareq}) has one solution with respect to $\bar s$:
\begin{equation}
\bar s (s) = - {D_1(s) \over \alpha \mu_* D_*(s)} =
{D_1 (s) \over \alpha |\mu_*| D_*(s)} \, .
\end{equation}

Again, it can be verified that the equation $\bar s (s) = s$ has exactly
one solution in the range $s > 0$ if $\bar s'(0) > 1$, and there are no
solutions in the opposite case $\bar s' (0) \le 1$.  Calculating the derivative
$\bar s' (0)$, we obtain that exactly one tachyonic mass eigenvalue is present
in the theory if
\begin{equation}
|\mu_*| < \frac12 \left(\alpha^2 - 1 \right) \, ,
\end{equation}
and tachyonic modes are absent if the value of the Planck mass $m_*$ is
sufficiently large, namely, if
\begin{equation}\label{masstar}
|\mu_*| \ge \frac12 \left(\alpha^2 - 1 \right) \, .
\end{equation}

We note that our method of finding the range of parameters where tachyonic
terms are present or absent is not restricted to the model under investigation
and can be used whenever the equation for tachyonic modes has the form
(\ref{E}), as is the case, e.g., in the theory with arbitrary signs of the
brane gravitational constants \cite{CD,Padilla,Smolyakov,mirror} and/or with
the Gauss--Bonnet action in the bulk \cite{CD}.

The issue of ghosts in the gravitational sector of the complementary theory
with positive value of $M$ but arbitrary signs of the brane gravitational
constants was considered in \cite{Padilla}, and we apply the results obtained
therein to our case. First, we start with the radion. The radion degree of
freedom in our formalism is connected with the possibility of brane bending in
the bulk. After identifying the physical degrees of freedom for the radion, one
can obtain the conditions for its ghost-free character in our model using the
results of \cite{Padilla}:
\begin{equation}\label{ghost-free}
M^3 \left({1 \over 1 - 2 \mu_*} - {e^{- 2 k \rho} \over 1 + 2 \mu} \right) \geq
0 \, ,
\end{equation}
which we expressed in terms of our parameters $\mu = k m^2 / M^3$ and $\mu_* =
k m_*^2 / M^3$ restricting ourselves to the case $k > 0$ and taking into
account that $M$ can be of any sign. Then the conditions of absence of both
tachyons and radion ghosts in the case $M < 0$ following from (\ref{mass}),
(\ref{masstar}), and (\ref{ghost-free}) are
\begin{equation}\label{tachyghost}
\mu_* = 0 \, , \ \ \frac12 \left(1 - e^{- 2 k \rho} \right) \le |\mu| < \frac12
\, , \quad \mbox{and} \quad \mu = 0 \, , \ \ |\mu_*| \ge \frac12 \left(e^{2 k
\rho} - 1 \right) \, .
\end{equation}
These conditions on the constants of the theory can be seen to be rather
restrictive.

Following \cite{Padilla}, we can also show that the massive gravitational modes
in the theory under consideration have ghost-like nature. For free metric
perturbations in the form (\ref{induced}), (\ref{scale}) described by the
transverse traceless modes $\gamma_{\alpha\beta} (x, y)$ with the boundary
conditions (\ref{b-hidden}) and (\ref{b-visible}) in which we set $\phi = 0$
and $\tau_{\alpha\beta} = 0$, one obtains the gravitational part of action
(\ref{action}) to quadratic order in the form
\begin{eqnarray}
S &=& {M^3 \over 2} \int_0^\rho dy e^{- 2ky} \int dx \left(
\gamma^{\alpha\beta} \Box \gamma_{\alpha\beta} - e^{- 2ky} \partial_y
\gamma^{\alpha\beta} \partial_y \gamma_{\alpha\beta} \right) \nonumber \\ &+&
{m^2 \over 2} \int\limits_{y = 0} dx  \gamma^{\alpha\beta} \Box
\gamma_{\alpha\beta} + {m_*^2 \over 2} e^{- 2 k\rho} \int\limits_{y = \rho} dx
\gamma^{\alpha\beta} \Box \gamma_{\alpha\beta} \, . \label{secord}
\end{eqnarray}
Expanding the perturbation in the modes $\psi(q,y)$ that are solutions of
Eq.~(\ref{eq-psi}) with the corresponding boundary conditions,
\begin{equation}
\gamma_{\alpha\beta} (x, y) = \sum_q \chi_{\alpha\beta} (q, x) \psi (q, y) \, ,
\end{equation}
substituting this expansion into action (\ref{secord}), and using the
orthogonality condition
\begin{equation}
\displaystyle M^3 \int_0^\rho dy e^{-2ky} \psi (q_1,y) \psi (q_2, y) + m^2 \psi
(q_1,0) \psi (q_2, 0) + m_*^2 e^{-2k\rho} \psi (q_1, \rho) \psi (q_2, \rho) = 0
\end{equation}
for $q_1 \ne q_2$, we arrive at the following quadratic effective action (cf.\@
with \cite{Padilla}):
\begin{equation}
S = \frac12 \sum_q C_q \int dx \chi^{\alpha\beta} (q, x) \left( \Box - q^2
\right) \chi_{\alpha\beta} (q, x) \, ,
\end{equation}
where
\begin{equation}
C_q = M^3 \int_0^\rho dy e^{-2ky} [\psi (q, y)]^2 + m^2 [\psi (q, 0)]^2 + m_*^2
e^{-2k\rho} [\psi (q, \rho)]^2 \, .
\end{equation}
For the massless mode ($q = 0$), we have $\psi(0, y) \equiv {\rm const}$, and
the constant $C_0$ is given by
\begin{equation}
C_0 = {M^3 \over 2k} [\psi (0, 0)]^2 \left[ 1 + 2 \mu - \left(1 - 2 \mu_*
\right) e^{-2k\rho} \right]
\end{equation}
and is positive in all cases in which tachyonic modes are absent in the theory,
as can be seen from conditions (\ref{mass}) and (\ref{masstar}).  Thus, the
massless graviton is not a ghost.  However, for the massive modes ($q \ne 0$),
using Eq.~(\ref{eq-psi}),  one can obtain the expression
\begin{equation}
C_q = {M^3 \over q^2} \int_0^\rho dy e^{-4ky} [\psi' (q, y)]^2 \, , \quad q \ne
0 \, ,
\end{equation}
which shows that the massive modes have positive norm in the case $M > 0$, and
have ghost-like nature in the case $M < 0$.

\section{Corrections to Newton's law} \label{corr}

In this section, we compute the gravitational potential $V(r)$ on the visible
brane induced by a static point source located on the visible or hidden brane
and determine corrections to the Newton's law in the physically reasonable case
$k > 0$, i.e., where the zero-mode graviton is localized around the visible
brane.  Our starting formula is
\begin{equation}
h^{\rm (vis)}_{00} = - (1 + 2 V) \, ,
\end{equation}
where $h^{\rm (vis)}_{\alpha\beta}$ is the induced metric on the visible brane.

\subsection{Matter source on the visible brane}

If matter source is on the visible brane only, then the induced metric $h^{\rm
(vis)}_{\alpha\beta}$ is given by Eq.~(\ref{indmet}). The stress--energy tensor
of a static point source of mass ${\cal M}$ is $\tau_{00} = {\cal M} \delta
\left( {\vec r\,} \right)$ with other components being zero, and $\tau = -
\tau_{00}$. Its Fourier image is $\tau_{00} (p_\alpha) = 2 \pi {\cal M} \delta
\left(p_0\right)$, containing only tachyonic modes.  Hence, we use the formulas
of Sec.~\ref{tachyon}.

We take into account (\ref{radion}) and (\ref{psi-t}) for the Fourier transform
of the solution for the induced metric on the first brane with the source on
the same brane.  Collecting all expressions together, we obtain the Fourier
image of the gravitational potential
\begin{equation} \label{vp}
V (p_\alpha) = {2 \pi {\cal M} \delta \left( p_0 \right) \over 3} \left[ {k
\over A {\vec p\,}^2} - {2 \over M^3 |{\vec p\,}|} \cdot f \Bigl( |{\vec p\,}|
\Bigr) \right] \, ,
\end{equation}
where the function $f (p)$ denotes the second fraction in expression
(\ref{psi-t}) taken at the position of the visible brane ($z = z_0$):
\begin{equation}
f(p) \equiv {C_K^* I_2 \Bigl( |z_0| \Bigr) + C_I^* K_2 \Bigl( |z_0| \Bigr)
\over C_I^* C_K^0 - C_I^0 C_K^*} \, .
\end{equation}
The potential $V(r)$ is obtained by taking the inverse Fourier transform of
(\ref{vp}):
\begin{equation}
V(r) = - {k {\cal M} \over 3 \pi^2 M^3 r} \left[ I (r) - {\pi M^3 \over 4 A}
\right] \, ,
\end{equation}
where
\begin{equation} \label{I}
I(r) = \int\limits_0^\infty d s \sin ( k r s )\, \Psi (s) \, , \qquad \Psi(s) =
{D (s) + \mu_* \alpha s D_2 (s) \over E(s) } \, ,
\end{equation}
and $E(s)$ is given by Eq.~(\ref{E}).

The integral in (\ref{I}) cannot be evaluated exactly, but it can be
approximated in different regions of $r$, as it is done, e.g.,  in
\cite{KTT,JKP}.

{\bf 1.} On very large spatial scales $kr \gg \alpha \equiv e^{k\rho}$, we need
the asymptotics of the function $\Psi (s)$ in the region $\alpha s \ll krs \sim
1$.  It is given by the expression
\begin{eqnarray}
\Psi(s) &\sim& {2 \over 1 + 2 \mu + \alpha^{-2} \left(2 \mu_* - 1 \right) }
\cdot \frac{1}{s} \nonumber
\\  &+& { \left( \alpha^2 - 1 \right) (2 \mu_* - 1) \left[ 3 \alpha^2 - 1 + 2
\mu_* \left( 1 - \alpha^2 \right) \right] + 4 \alpha^2 \ln\alpha \over
 4 \left[ \alpha^2 (1 + 2 \mu) + 2 \mu_* - 1 \right]^2} \cdot s \, ,
\end{eqnarray}
and the integral in (\ref{I}) is approximated by using the regularization
\cite{KTT,JKP}
\begin{equation}
\int\limits_0^\infty s \sin (krs) ds \quad \rightarrow \quad \lim_{\epsilon \to
0} \int\limits_0^\infty s \sin (krs) e^{- \epsilon s} ds = 0
\end{equation}
with the result
\begin{equation}
I (r) \approx {\pi \over 1 + 2 \mu + \alpha^{-2} \left(2 \mu_* - 1 \right) } =
{\rm const} \, .
\end{equation}
The potential in this region has Newtonian form
\begin{equation}
V(r) = - {G {\cal M} \over r} \, , \qquad G = G_{\rm N} \left(1 + {B \over 3 A
\alpha^2 } \right) \, ,
\end{equation}
where $G_{\rm N}$ given by (\ref{newton}), which is in complete agreement with
the zero-mode approximation (\ref{g-vis}).  The theory has continuous Newtonian
limit as $\alpha \to \infty$.

{\bf 2.} On the scales $1 \ll kr \ll \alpha$, it is the region of integration
$1/\alpha \ll s \ll 1$ that substantially contributes to the integral in
(\ref{I}). In this region, we have
\begin{equation}
\Psi (s) \approx {1 \over 1 + 2 \mu} \cdot \frac1s - {\log (s/2) \over (1 + 2
\mu)^2 } \cdot s \, ,
\end{equation}
substituting which to (\ref{I}), we obtain
\begin{equation}
I (r) = {\pi \over 1 + 2 \mu} + {\pi \over 2 (1 + 2 \mu )^2} \cdot {1 \over
(kr)^2}
\end{equation}
and
\begin{equation}
V(r) = - {G {\cal M} \over r} \left( 1 + {2 M^3 \over 3 A (kr)^2} \right) \, ,
\qquad G = G_{\rm N} \left( 1 - {B \over A \alpha^2} \right) \, .
\end{equation}

We observe corrections to the Newtonian potential similar to those of the
Randall--Sundrum model \cite{RS2} but with somewhat different relative
constant.

{\bf 3.} In the case $kr \ll 1$, we can use the asymptotics for the function
$\Psi (s)$ at infinity:
\begin{equation}
\Psi (s) \sim {1 \over \mu s + 1 + 15 \mu / 8} \left[1 + {\cal O} \left( s^{-1}
\right) \right] \, , \quad \mu \ne 0 \, .
\end{equation}
This case is further partitioned into two asymptotic regions, depending on the
magnitude of $\mu$.

3a. $kr \ll |\mu|$.  Here, if $\mu$ is negative, then it cannot be small by
absolute value since, in this latter case, the theory contains tachyons. We
obtain
\begin{equation}
I (r) \approx {\pi \over 2 \mu} + \frac1\mu \left({15 \over 8} + \frac1\mu
\right) kr \log \left[\left({15 \over 8} + \frac1\mu \right) kr \right]
\end{equation}
and
\begin{equation}
V (r) = - {G {\cal M} \over r}  - {k {\cal M} \over 3 \pi^2 m^2} \left({15
\over 8} + \frac1\mu \right) \log \left[\left({15 \over 8} + \frac1\mu \right)
kr \right] \, , \qquad G = {1 \over 8 \pi m^2 } \cdot {\mu + 2/3 \over \mu +
1/2} \, .
\end{equation}
The logarithmic corrections in these expressions assume that the expression
$15/8 + 1/ \mu $ is not very small by absolute value.

3b. $\mu \ll kr \ll 1$.  Here we must consider only positive $\mu$.  In this
case,
\begin{equation}
I (r) \approx {1 \over kr} \, ,
\end{equation}
and the gravitational law is five-dimensional:
\begin{equation}
V (r) = - {{\cal M} \over 3 \pi^2 M^3 r^2} \left( 1 - \frac\pi4 kr \right) \, .
\end{equation}

The expressions for the gravitational potentials obtained in cases 3a and 3b
are analogous to those of \cite{DGP}.  The same results would be obtained in
the linear approximation in the one-brane case considered in \cite{Tanaka}, as
is clear from the remark made after Eq.~(\ref{Cs}), which identifies the
corresponding propagators.

\subsection{Mater source on the hidden brane}

In a similar way one can consider the case where the stationary matter resides
on the hidden brane with mass ${\cal M}_*$ defined as $\tau^*_{00} = {\cal M}_*
\delta ({\vec r\,})$. In this case, we obtain the following expression for the
gravitational potential on the visible brane:
\begin{equation}
V(r) = - {k {\cal M}_* \over 3 \pi^2 M^3 \alpha r } I(r) \, ,
\end{equation}
where
\begin{equation} \label{I2}
I(r) = \int\limits_0^\infty ds \sin (krs) \Psi (s) \, , \qquad \Psi(s) = {I_1
(s) K_2 (s) + I_2 (s) K_1 (s) \over E(s) } \, ,
\end{equation}
and $E(s)$ is given by Eq.~(\ref{E}).

The asymptotic expressions for $\Psi (s)$ can be found in various regions:
\begin{equation}
\Psi (s) \approx  {2 \alpha \over 2 \mu_* - 1 + \alpha^2 ( 2 \mu + 1) } \cdot
\frac1s + {\cal O} (s) \, , \quad s \ll \alpha^{-1} \, ,
\end{equation}
\begin{equation}
\Psi(s) \approx {\sqrt{2 \pi} e^{- \alpha s} \over (1 + 2 \mu) \mu_*
\sqrt{\alpha s}} \left[1 + \left( \frac{15}{8} - \frac1{\mu_*} \right) {1 \over
\alpha s} + {\cal O} (s^2) \right] \, , \quad \alpha^{-1} \ll s \ll 1 \, ,
\quad \mu_* \ne 0 \, ,
\end{equation}
\begin{equation}
\Psi(s) \approx {\sqrt{2 \pi \alpha s}\, e^{- \alpha s} \over (1 + 2 \mu) }
\left[1 + \frac{3}{8\alpha s} + {\cal O} (s^2) \right] \, , \quad \alpha^{-1}
\ll s \ll 1 \, , \quad \mu_* = 0 \, ,
\end{equation}
\begin{equation}
\Psi (s) \approx {2 \sqrt\alpha e^{- \alpha s} \over \mu \mu_* \alpha s^2 +
\Bigl( \mu + \mu_* \alpha + 15 \mu \mu_* \alpha / 8 \Bigr) s} \, , \quad s \gg
1 \, , \quad \mu^2 + \mu_*^2 \ne 0 \, ,
\end{equation}
\begin{equation}
\Psi (s) \approx 2 \sqrt\alpha\, e^{- \alpha s} \, , \quad s \gg 1 \, , \quad
\mu = \mu_* = 0 \, .
\end{equation}

Using these expression, it is not difficult to obtain the estimates for the
gravitational potential $V(r)$ caused by the presence of the static source on
the hidden brane in various regions.  We have
\begin{equation}
V(r) \approx - {4 G_{\rm N} {\cal M}_* \over 3 \alpha^2 r} \, , \quad kr \gg
\alpha \, ,
\end{equation}
where $G_{\rm N}$ is given by (\ref{newton}).  Again, in this distance range,
the result can be obtained by using the zero-mode approximation (\ref{g-hid}).

In the case $kr \ll \alpha$, the result crucially depends on whether $\mu_*$ is
zero or not:
\begin{equation} \label{mu0}
V(r) \approx - {c_1 k^2 {\cal M}_* \over \pi^2 A \alpha^3 } \left[ 1 - c_2
\left( {kr \over \alpha } \right)^2 \right] \, , \quad kr \ll \alpha \, , \quad
\mu_* = 0 \, ,
\end{equation}
\begin{equation} \label{mu}
V (r) \approx - {2 k^2 {\cal M}_* \over 3 \pi^2 A \alpha^3 } \left[ \left( 1 +
{c_3 \over \mu_* } - {c_4 \over \mu_*^2} \right) - \left( c_5 + {c_6 \over
\mu_* } - {c_7 \over \mu_*^2} \right) \cdot \left( { kr \over \alpha }
\right)^2 \right] \, , \quad kr \ll \alpha \, , \quad \mu_* \ne 0 \, ,
\end{equation}
Here the constants $c_n$ take the following approximate values:
\begin{equation}
c_1 \approx 1.77 \, , \quad c_2 \approx 1.02 \, , \quad c_3 \approx 1.3 \, ,
\quad c_4 \approx 0.35 \, , \quad c_5 \approx 0.06 \, , \quad c_6 \approx 1.12
\, , \quad c_7 \approx 0.24 \, .
\end{equation}
Expression (\ref{mu}) is not valid for sufficiently small $\mu_*$ since we know
that, in the limit of $\mu_* \to 0$, the asymptotics changes to (\ref{mu0}). In
fact, comparison with the exact numerical integration of (\ref{I2}) shows that
our approximate result (\ref{mu}) is only good for $|\mu_*| \gsim 0.5$.

\section{Discussion}

It is known that the braneworld model becomes rather rich in its cosmological
manifestations if curvature term is present in the action for the brane (see
\cite{DGP,CHS,DDG,KTT,SS,AS,SS1}).  In this paper, we studied the model with
scalar-curvature terms for the branes on the original Randall--Sundrum
two-brane background. The linearized gravitational equations (\ref{g-vis}) and
(\ref{g-hid}) in this case have the same structure as in the original
Randall--Sundrum model but with different physical constants.  In the limit of
vanishing brane Planck masses $m$ and $m_*$, they tend to the known results
\cite{KTT,SV,JKP}, which are physically reasonable only for the bulk Planck
mass $M > 0$. In the opposite limit of $M \to 0$ while $k = \sigma/M^3$ is
fixed, they produce reasonable results (\ref{lim-vis}) and (\ref{lim-hid})
independently of the sign of $M$.

In this paper, we developed a general method for detecting tachyonic modes in
the braneworld theory, which can be generalized to theories giving rise to
equations of the similar kind, e.\,g., the theory with Gauss--Bonnet term in
the bulk action \cite{CD}. In our case, for negative values of $M$, the
linearized theory typically contains tachyonic modes in the gravitational
sector.  If both brane Planck masses are nonzero, then we have one or two
tachyonic mass eigenvalues depending on the constants of the theory, the
conditions of which were determined in Sec.~\ref{tachyon}. However, in the case
where one of the brane Planck masses is zero, tachyonic modes are absent if the
other brane has sufficiently high Planck mass [given by Eq.~(\ref{mass}) for
the visible brane, and by Eq.~(\ref{masstar}) for the hidden brane].  In the
case of negative bulk Planck mass $M$, the zero-mode graviton is `localized'
around the brane with {\em negative\/} tension and is not a ghost in all cases
where tachyonic modes are absent in the theory. However, in all cases with
negative $M$, the massive gravitational modes of the theory under consideration
have ghost-like character.  The conditions of absence of both radion ghosts and
gravitational tachyons are expressed by (\ref{tachyghost}).

Exploring both signs of the bulk gravitational constant may be interesting in
connection with some braneworld cosmological models requiring {\em negative\/}
brane tension, such as the model of {\em disappearing dark energy\/} (DDE)
recently discussed in \cite{SS,AS} or the braneworld models with `quiescent'
cosmological singularities during expansion \cite{SS1}. The DDE model
\cite{SS,AS} represents a cosmological braneworld with the Randall--Sundrum
constraint (\ref{lambda-rs}), negative brane tension, and the condition $|\mu|
\ge 1/2$, which is required for physical consistency and which implies
inequality (\ref{mass}).  The `quiescent' singularities in the AdS-embedded
braneworld models occur during the universe expansion and are characterized by
{\em finiteness\/} of the scale factor, Hubble parameter, and matter density.
The braneworld cosmological equations involve the bulk gravitational constant
only in even power; therefore, their behaviour is independent of its sign on
the homogeneous and isotropic level \cite{SS}. However, as noted in the
introduction, the sign of the bulk gravitational constant is of importance for
the small-scale gravitational physics in a braneworld universe, in particular,
for the behaviour of cosmological perturbations. The results of the present
paper indicate that models with negative bulk gravitational constant can be
free from tachyons, although they are plagued with massive ghosts in the
gravitational sector. Perhaps, the unwanted situation with ghosts can be
remedied by modifications of the bulk action. It should be emphasized that
these results do not relate to the braneworld cosmological models with negative
brane tension but positive bulk and brane gravitational constants, which
require future investigation.

\section*{Acknowledgments}

The authors are grateful to Varun Sahni for valuable comments and suggestions.
Yu.~S. acknowledges warm hospitality of the Inter-University Centre for
Astronomy and Astrophysics (IUCAA) in Pune, India.

\end{document}